\documentclass[a4paper]{interact}

\usepackage{color}

\usepackage{epstopdf}
\usepackage[caption=false]{subfig}

\usepackage[numbers,sort&compress]{natbib}
\bibpunct[, ]{[}{]}{,}{n}{,}{,}
\renewcommand\bibfont{\fontsize{10}{12}\selectfont}
\makeatletter
\def\NAT@def@citea{\def\@citea{\NAT@separator}}
\makeatother

\theoremstyle{plain}

\theoremstyle{definition}

\theoremstyle{remark}

\usepackage{braket}

\begin{document}

\articletype{ARTICLE TEMPLATE}

\title{Bayesian inference to identify crystalline structures for XRD}

\author{
    \name{
    Ryo Murakami\textsuperscript{a}, Yoshitaka Matsushita\textsuperscript{a}, Kenji Nagata\textsuperscript{a}, Hayaru Shouno\textsuperscript{b}, Hideki Yoshikawa\textsuperscript{a}
    }
    \affil{
        \textsuperscript{a}National Institute for Materials Science, Tsukuba 305-0044, Japan,\\
        \textsuperscript{b}The University of Electro-Communications, Chofu, 182-8585, Japan,
    }
}

\maketitle

\begin{abstract}
Crystalline phase structure is essential for understanding the performance and properties of a material. Therefore, this study identified and quantified the crystalline phase structure of a sample based on the diffraction pattern observed when the crystalline sample was irradiated with electromagnetic waves such as X-rays. Conventional analysis necessitates experienced and knowledgeable researchers to shorten the list from many candidate crystalline phase structures. However, the Conventional diffraction pattern analysis is highly analyst-dependent and not objective. Additionally, there is no established method for discussing the confidence intervals of the analysis results. Thus, this study aimed to establish a method for automatically inferring crystalline phase structures from diffraction patterns using Bayesian inference. Our method successfully identified true crystalline phase structures with a high probability from 50 candidate crystalline phase structures. Further, the mixing ratios of selected crystalline phase structures were estimated with a high degree of accuracy. This study provided reasonable results for well-crystallized samples that clearly identified the crystalline phase structures. 
\end{abstract}


\section{Introduction}\label{sec:introduction}
Crystalline phase structure is essential for understanding the performance and properties of a material. Therefore, this study identified and quantified the crystalline phase structure of a sample based on the diffraction pattern observed when the crystalline sample was irradiated with electromagnetic waves such as X-rays. The measurement of the diffraction patterns using X-rays as probes is known as X-ray diffraction (XRD). The crystal structure of a material can be understood by analyzing the diffraction peaks in the XRD data.

A typical XRD data analysis method involves a simple comparison of the measured XRD data with a database. This method first detects the diffraction peaks in the measured XRD data by the smoothed derivative\cite{1990savitzky,1973peakDetermination,1988peakDetermination}. Thereafter, the diffraction angles of the detected peaks are compared with those of the diffraction patterns registered in the database and the similarity to the diffraction patterns in the database is calculated. The diffraction patterns ranked by similarity are suggested by an analyst. Thus, in a typical analysis, the experience and knowledge of the researcher are crucial to shorten the list from several candidate crystal structures. However, the typical diffraction pattern analysis is highly analyst-dependent and not objective. Additionally, there is no established method for discussing the confidence intervals of the analysis results. Consequently, the interpretation of the analysis results is highly dependent on the analysts. Diffraction pattern analysis methods have been proposed to solve such analytical problems.

In recent years, methods for diffraction-pattern analysis using Bayesian estimation have been proposed, allowing confidence intervals to be discussed\cite{2016UseBayesian}. In addition, black-box optimization methods have been proposed for hyperparameters that are subjectively determined by an analyst\cite{2022AutomatedData}. The proposed method is effective for solving several problems in diffraction pattern analyses. However, this has not been sufficiently discussed from the perspective of automatic estimation of the crystal structure contained in a measured sample from the diffraction pattern. Identifying the crystalline phase structures contained in a diffraction pattern is challenging because the number of candidate crystalline phase structures can be in the order of tens or hundreds, leading to combination explosions. Moreover, this problem requires considerable computational time because the crystal structure contains dozens of diffraction peaks. Despite these challenges, it is necessary to establish a method for identifying crystalline phase structures from diffraction patterns with confidence intervals (probability).

This study aimed to establish a method for the automatic estimation of crystalline phase structures from diffraction patterns. The proposed method decomposes the measured diffraction patterns and automatically selects crystalline phase structures using the diffraction patterns measured at each institute associated with the crystal structures or obtained via simulations as basis functions. The proposed method makes three main contributions to literature.
\begin{itemize}
    \item Crystalline phase structures can be selected precisely and automatically.
    \item Posterior distributions can be estimated (confidence intervals can be discussed).
    \item A global solution is provided (no initial value dependence).
\end{itemize}
The proposed method, which extracts material descriptors corresponding to the crystal structure from measured diffraction patterns, is expected to play an important role in promoting the development of data-driven materials. Note that in this paper the term "crystal structure" refers specifically to the crystalline phase structure.


\begin{figure}
    \centering
    \includegraphics[width=\linewidth]{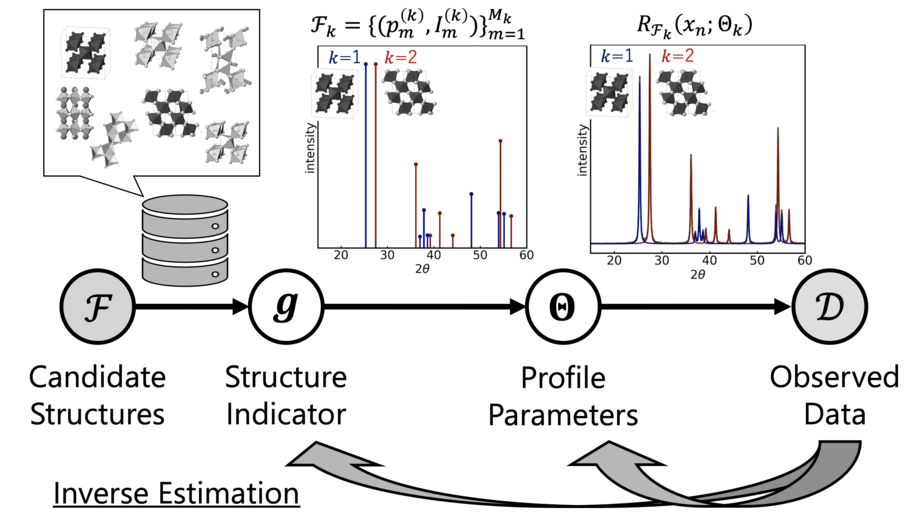}
    \caption{Observation process of XRD data and a conceptual diagram of the proposed method, that is, Bayesian inverse estimation to identify the crystalline phase structure and their known structures for XRD analysis\cite{vista}.}
    \label{img:method:graphical-abstract}
\end{figure}

\section{Concept}
\quad Figure \ref{img:method:graphical-abstract} shows an observation process of XRD data and a conceptual diagram of the proposed method. We suppose a multitude of candidate crystal phases and structures $\mathcal{F}$ when preparing the materials. The crystal structures contained in the material are selected by material synthesis, manufacturing processes, etc. This study treats the control variable dealing with crystal structure selection as the indicator variable $\bm{g} \in \{0, 1\}$. Ideally, the crystalline materials produced should have diffraction line spectra corresponding to the crystal phases and structures they contain. In practice, we observe diffraction peaks whose shapes are dependent on the profile parameters $\bm{\Theta}$ that correspond to the measurement environment. We considered a situation wherein only the observed diffraction data $\mathcal{D}$ and candidate crystal structures $\mathcal{F}$ were provided.

This study aimed to inversely estimate the structural indicator $\bm{g}$ and profile parameter set $\bm{\Theta}$ from the observed diffraction data (XRD data) shown in Figure \ref{img:method:graphical-abstract}. The proposed method is a Bayesian inverse estimation method used to identify crystal structures for XRD analysis.

\section{Model}\label{sec:model}
\subsection{Problem setting}
\quad The purpose is to estimate the profile parameters and the crystalline phase structures in the measured sample, considering the measured XRD data $\mathcal{D}=\{(x_i, y_i)\}_{i=1}^{N}$ and the candidate crystal structure $\mathcal{F}$. Here, $x_i \in (0, 180)$ and $y_i \in \mathbb{N}$ denote the diffraction angle $2\theta$ [$^{\circ}$] and the diffraction intensity [counts], respectively.

The candidate crystal structure factor set $\mathcal{F}$ is expressed as:
\begin{eqnarray}
    \mathcal{F} &=& \{\mathcal{F}_{k} \ | \ k \in \{1, 2, ..., K\}\},\\
    \text{where} \:\: \mathcal{F}_{k} &=& \{(p^{(k)}_{m}, I^{(k)}_{m}) \ | \ m \in \{1, 2, ..., M_k\}\} \subset \mathcal{F},
\end{eqnarray}
where $K\in\mathbb{N}$ is the number of candidate crystal structures and $\mathcal{F}_k$ is the $k$-th crystal structure factor. The elements of the crystal structure factor $p^{(k)}_{m} \in (0, 180)$ and $I^{(k)}_{m} \in [0, 1]$ are the diffraction angle (peak position) [$^{\circ}$] and relative intensity of the $m$-th diffraction peak in $\mathcal{F}_k$ for a crystal structure $k$. Further, $M_{k}\in\mathbb{N}$ denotes the number of peaks in $\mathcal{F}_k$. In this study, the candidate crystal structure factor set $\mathcal{F}$ is provided.


\subsection{Profile function}
\quad XRD data can be represented by a profile function $f_{\mathcal{F}}(x_i;\Theta):\mathbb{R} \rightarrow \mathbb{R}^{+}_{0}$, which is a linear sum of the signal spectrum $S_{\mathcal{F}}(x_i;\Theta_\mathrm{S})$ and the background $B(x_i;\Theta_\mathrm{B})$:
\begin{eqnarray}
        y_i &\approx& f_{\mathcal{F}}(x_i;\Theta),\\
        &=& S_{\mathcal{F}}(x_i;\Theta_\mathrm{S}) + B(x_i;\Theta_\mathrm{B}),
\end{eqnarray}
where $(x_i, y_i)$ denote the measured data points, the function $S_{\mathcal{F}}(x_i;\Theta_\mathrm{S})$ denotes the signal spectrum based on the candidate crystal structures $\mathcal{F}$, and the function $B(x_i;\Theta_\mathrm{B})$ denotes the background. We set $\Theta=\{\Theta_\mathrm{S}, \Theta_\mathrm{B}\}$ as the profile parameter set. In addition, the sets $\Theta_\mathrm{S}$ and $\Theta_\mathrm{B}$ are the signal spectrum and background parameter sets, respectively.


The signal spectrum $S_{\mathcal{F}}(x_i;\Theta_\mathrm{S})$ is expressed as a linear sum of the profile function (peaks) $C_{\mathcal{F}_{k}}(x_i;\Theta^{(k)}_\mathrm{S}):\mathbb{R} \rightarrow \mathbb{R}^{+}_{0}$ in a crystal structure $\mathcal{F}_{k}$ among the several candidates\cite{1990ProfileFunction}:
\begin{eqnarray}\label{eq:xrd:signal}
    S_{\mathcal{F}}(x_i;\Theta_\mathrm{S}) &=& \sum_{k=1}^{K}{
        h_{k} C_{\mathcal{F}_{k}}(x_i;\Theta^{(k)}_\mathrm{S})
    },
\end{eqnarray}
where $h_k \in \mathbb{R}^{+}$ denotes the signal intensity of crystal structure factor $\mathcal{F}_k$. The profile function $C_{\mathcal{F}_{k}}(x_i;\Theta^{(k)}_\mathrm{S})$ of candidate crystal structure $k$ is defined as follows:
\begin{eqnarray}
    C_{\mathcal{F}_{k}}(x_i;\Theta^{(k)}_\mathrm{S}) &=& \sum_{m=1}^{M_{k}}{
        I_{m}^{(k)}
        V \left (
            x_i;
            \rho_{mk},
            \Sigma_k,
            \Omega_k,
            r_k
        \right )
    },\\
    &=& \sum_{m=1}^{M_{k}}{
        I_{m}^{(k)} \{(1-r_k) G(x_i;\rho_{mk}, \Sigma_k) + r_k L(x_i;\rho_{mk}, \Omega_k)\}},\\
    \text{where} \:\: \rho_{mk} &=& p_{m}^{(k)}+\mu_k,
\end{eqnarray}
where $\mu_k \in \mathbb{R} and r_k \in [0, 1]$ are the peak shift and Gauss-Lorentz ratio at the peak of crystal structure $k$, respectively, $\rho_{mk} \in \mathbb{R}$ is the peak position of the peak function, and the function $V(x_i):\mathbb{R} \rightarrow \mathbb{R}^{+}_{0}$ is a pseudo-Voigt function\cite{1974PseudoVoigt}. In addition, $G(x_i):\mathbb{R} \rightarrow \mathbb{R}^{+}_{0}$ and $L(x_i):\mathbb{R} \rightarrow \mathbb{R}^{+}_{0}$ are Gaussian and Lorentz functions, respectively. $\Sigma_{k}=\Sigma(x_i;u_k, v_k, w_k, \alpha_{k}):(0, 180) \rightarrow \mathbb{R}^{+}$ and $\Omega_{k}=\Omega(x_i;s_k, t_k, \alpha_{k}):(0, 180) \rightarrow \mathbb{R}^{+}$ are the Gaussian and Lorentzian widths of the peak, respectively, as a function of the diffraction angle $x_i \in (0, 180)$. The width functions $\Sigma_{k}(x_i)$ and $\Omega_{k}(x_i)$ are expressed as
\begin{eqnarray}
    \Sigma(x_i;u_k, v_k, w_k, \alpha_{k}) &=& A(x_i;\alpha_{k}) \sqrt{ u_{k}\tan^{2}\left(\frac{x_i}{2}\right)-v_{k}\tan\left(\frac{x_i}{2}\right)+w_{k} },\\
    \Omega(x_i;s_k, t_k, \alpha_{k}) &=& A(x_i;\alpha_{k}) \left\{ s_{k}\sec\left(\frac{x_i}{2}\right)+t_{k}\tan\left(\frac{x_i}{2}\right) \right\},\\
    \text{where} \:\: A(x_i;\alpha_{k}) &=& \left\{
        \begin{array}{ll}
        \alpha_{k} & (x_{i} \geq \rho_{k})\\
        1 & (x_{i} < \rho_{k}),
        \end{array}
        \right.\\
        &=& \operatorname{sign}(x_{i}-\rho_{k})\frac{\alpha_{k} - 1}{2} + \frac{\alpha_{k} + 1}{2},
\end{eqnarray}
where $\{u_k, v_k, w_k\}$ and $\{s_k, t_k\}$ are the Gaussian and Lorentzian width parameter sets, respectively. Function $A(x_i;\alpha_{k}):\mathbb{R} \rightarrow \mathbb{R}$ is a function expressing the peak asymmetry, and $\alpha_{k} \in \mathbb{R}^{+}$ is the asymmetry parameter for the peak function. Further, the function $\mathrm{sign}(\cdot):\mathbb{R} \rightarrow \{-1, 1\}$ is the sign function and the trigonometric function $\sec(x)$ is $\sec(x)=1/\cos(x)$.

The optimization parameter set for the signal spectrum $S_{\mathcal{F}}(x_i;\Theta_\mathrm{S})$ is expressed as:
\begin{eqnarray*}
    \Theta_\mathrm{S} &=& \{\Theta^{(k)}_\mathrm{S} \ | \ k \in \{1, 2, ..., K\}\},\\
    \text{where} \:\: \Theta^{(k)}_\mathrm{S} &=& \{(h_k, \mu_k, \alpha_k, r_k, u_k, v_k, w_k, s_k, t_k)\}.
\end{eqnarray*}
Background $B(x_i;\Theta_\mathrm{B}):\mathbb{R} \rightarrow \mathbb{R}$ is defined as follows:
\begin{eqnarray}
    B(x_i;\Theta_\mathrm{B}) &=& aV(x_i;0.0, \sigma_{\mathrm{bg}}, \sigma_{\mathrm{bg}}, r_{\mathrm{bg}}) + b,
\end{eqnarray}
where the background parameter set is $\Theta_\mathrm{B} = \{a, \sigma_{\mathrm{bg}}, r_{\mathrm{bg}}, b\}$.

\subsection{Generation Model}
\quad We assume that the observed data $\{(Y, X)\}=\{(x_i,y_i)\}_{i=0}^{N}$ are stochastically distributed owing to statistical noise in the measurement. Next, we consider the joint distribution $P(Y, \Theta)$, which can be expanded to $P(Y, \Theta)=P(\Theta|Y)P(Y)$. Using Bayes' theorem to swap the orders of $Y$ and $\Theta$, we can expand $P(Y, \Theta)=P(Y|\Theta)P(\Theta)$. Hence, the posterior distribution $P(\Theta|Y)$ is expressed as:
\begin{eqnarray}
    P(\Theta|Y) = \frac{P(Y|\Theta)P(\Theta)}{P(Y)} \propto P(Y|\Theta)P(\Theta),
\end{eqnarray}
where $P(\Theta|Y)$ and $P(\Theta)$ are the posterior and prior distributions, respectively, in the Bayesian inference. Further, $P(Y|\Theta)$ is the conditional probability of $Y$ given the model parameter set $\Theta$, which is a probability distribution explained by error theory.

To derive $P(Y|\Theta)$, we consider the observation process of $\{(x_i, y_i)\}$ at the observation data points. Assuming that the observed data are independent of each other, the conditional probability of the observed data $\{(Y, X)\}$ can be expressed as:
\begin{eqnarray}
    P(Y|\Theta) = \prod_{i=0}^{N}{P(y_i|\Theta)}.
\end{eqnarray}
As XRD spectra are count data, the conditional probability $P(y_i|\Theta)$ of the intensity $y_i$ for the diffraction angle $x_i$ follows a Poisson distribution $\mathcal{P}(y_i|f_{\mathcal{F}}(x_i;\Theta))$:
\begin{eqnarray}
    P(y_i|\Theta) &=& \mathcal{P}(y_i|f_{\mathcal{F}}(x_i;\Theta))\\
    &=& \frac{f_{\mathcal{F}}(x_i;\Theta)^{y_i}\exp{(-f_{\mathcal{F}}(x_i;\Theta))}}{y_i!}.
\end{eqnarray}
The cost function $E(\Theta) \in \mathbb{R}$ is defined by the negative log-likelihood function $E(\Theta)=-\ln{P(Y|\Theta)}$ and is expressed as follows:
\begin{eqnarray}
    E(\Theta) &=& -\sum_{i=0}^{N}{\ln{P(y_i|\Theta)}},\\
    &=& -\sum_{i=0}^{N}{
        \left\{ y_{i}\ln{f_{\mathcal{F}}(x_i;\Theta)} - f_{\mathcal{F}}(x_i;\Theta) - \ln{y_i!} \right\}
    }.
\end{eqnarray}
Further, $P(\Theta|Y)$ is expressed using the cost function $E(\Theta)$ and the prior distribution $P(\Theta)$ as follows:
\begin{eqnarray}
    P(\Theta|Y) &\propto& P(Y|\Theta)P(\Theta),\\
    &=& \exp{\{\ln{P(Y|\Theta)}\}}P(\Theta),\\
    &=& \exp{\{-E(\Theta)\}}P(\Theta).
\end{eqnarray}

\subsection{Identification of crystalline phase structures}
\quad In the analysis of XRD spectra, the crystal structure contained in the measured sample is often unknown. Therefore, it is important to accurately estimate the true crystal structure contained in the candidate crystal structures $\mathcal{F}$. We introduce an indicator vector $\bm{g}=\{g_k \in \{0, 1\} \ | \ k \in \{1,2,...,K\}\}$, which controls the existence of the crystal structure factors in Equation \eqref{eq:xrd:signal}:
\begin{eqnarray}
    f_{\mathcal{F}}(x_i;\bm{g}, \Theta) &=& S_{\mathcal{F}}(x_i;\bm{g}, \Theta_\mathrm{S}) + B(x_i;\Theta_\mathrm{B}),\\
    S_{\mathcal{F}}(x_i;\bm{g}, \Theta_\mathrm{S}) &=& \sum_{k=1}^{K}{
        g_{k} h_{k} C_{\mathcal{F}_{k}}(x_i;\Theta^{(k)}_\mathrm{S})
    },
\end{eqnarray}
where $g_k=1$ indicates that the crystal structure factor $\mathcal{F}_k$ is present in the sample. Conversely, $g_k=0$ implies that it is absent.

We now consider the joint distribution $P(\bm{g}, Y, \theta)$. $P(\bm{g}, Y, \theta)$ can be expanded to $P(\bm{g}, Y, \theta) = P(Y|\bm{g}, \Theta)P(\bm{g})P(\Theta)$. According to Bayes' theorem, the posterior distribution $P(\bm{g}, \Theta|Y)$ is expressed as:
\begin{eqnarray}
    P(\bm{g}, \Theta|Y) &\propto& P(Y|\bm{g}, \Theta)P(\bm{g})P(\Theta),\\
    &=& \exp{(-E(\bm{g}, \Theta))}P(\bm{g})P(\Theta).
\end{eqnarray}
The cost function $E(\bm{g}, \Theta)$ that introduces the indicator vector $\bm{g}$ is expressed as:
\begin{eqnarray}
    E(\bm{g}, \Theta) &=& -\sum_{i=0}^{N}{\ln{P(y_i|\bm{g}, \Theta)}},\\
    &=& -\sum_{i=0}^{N}{
        \left\{ y_{i}\ln{f_{\mathcal{F}}(x_i;\bm{g}, \Theta)} - f_{\mathcal{F}}(x_i;\bm{g}, \Theta) - \ln{y_i!} \right\}
    }.
\end{eqnarray}
Using the joint distribution presented above, the indicator vector $\bm{g}$ is estimated from the marginal posterior distribution as follows:
\begin{eqnarray}
    P(\bm{g}|Y) &=& \int{d\Theta P(\bm{g}, \Theta|Y)},\\
                &=& P(\bm{g})\int{d\Theta \exp{(-E(\bm{g}, \Theta))}P(\Theta)},
\end{eqnarray}
We estimate the profile and background parameters using the posterior distribution $P(\Theta|Y, \bm{g})$ on parameter set $\Theta$.

\section{Algorithm}\label{sec:algorithm}
\subsection{Replica Exchange Monte Carlo method | REMC method}
\quad We perform posterior visualization and the maximum a posteriori (MAP) estimation through sampling from the posterior distribution. A popular sampling method is the Monte Carlo (MC) method, which may be bounded by local solutions for cases when the initial value is affected or the cost function landscape is complex.

Therefore, the replica exchange Monte Carlo (REMC) method\cite{remc,2012BayesianSpectral} was used to estimate the global solution. For sampling using the REMC method, a replica was prepared with the inverse temperature $\beta$ introduced as follows:
\begin{eqnarray}
    P(\bm{g}, \Theta|Y;\beta=\beta_{\tau}) &=& \exp{(-\beta_{\tau} E(\bm{g}, \Theta))}P(\bm{g})P(\Theta),
\end{eqnarray}
where the inverse temperature $\beta$ is $0 = \beta_1 < \beta_2 < \cdots < \beta_{\tau} < \beta_T = 1$. For each replica, the parameters were sampled using the Monte Carlo method.

\section{Technique}\label{sec:technique}
\subsection{Tricks for high speeds}\label{subsec:tricks}
\quad This sub-section describes the techniques used to realize Bayesian inference of XRD spectra. In XRD spectral analysis, the number of candidate crystal structure factors $\{\mathcal{F}_k\}_{k=1}^{K}$ and the number of peaks $M_k$ for each crystal structure factor $\mathcal{F}_k$ are enormous. Therefore, to calculate the cost function $E()$ for each sample, multiple loops of $\sum^{N}_{i=1}{ \sum^{K}_{k=1}{ \sum_{m=1}^{M_k}{E(x_i)} } }$ must be computed. $M_k$ is an immutable value because it is inherently determined by the crystal structure. Although reduction in the number of data $N$ by downsampling is feasible, it is expected that the peak structure will be broken or the separation accuracy will be significantly reduced owing to sharp XRD peaks.

Herein, we focused on the number of candidate crystal structures $K$. This study screened the candidate crystallographic structure factors. To calculate the cost function $E()$, the crystal structures with $g_k=0$ need not be calculated. In other words, only the selected crystal structures $\{\mathcal{F}_k|g_k=1\}$ need to be considered. In the proposed method, we compute $\sum^{N}_{i=1}{\sum_{k\in\{\mathcal{F}_k|g_k=1\}}{\sum_{m=1}^{M_k}{E(x_i)}}}$ where $n(\{\mathcal{F}_k|g_k=1\})<K$.



\begin{figure}
    \centering
    \includegraphics[width=0.9\linewidth]{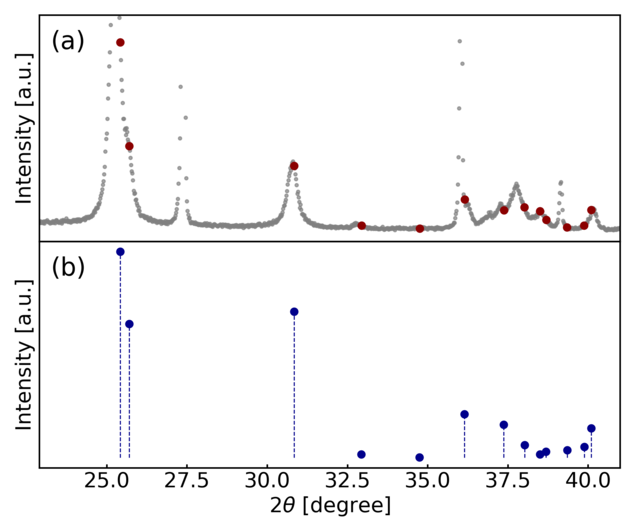}
    \caption{Supplementary diagram of the similarity calculation procedure.[(a) Observed XRD data $\mathcal{D}$ and (b) crystal structure factor $\mathcal{F}_k$].}
    \label{img:method:screening-overview}
\end{figure}

\subsection{Rough pre-screening}
\quad In this study, we screened candidate crystallographic structures as described in sub-section \ref{subsec:tricks}. This subsection describes the screening procedure. The similarity between the observed XRD data $\mathcal{D}$ and the crystal structure factor $\mathcal{F}_k$ was calculated and screening was performed by thresholding the similarity.

Figure \ref{img:method:screening-overview} presents a supplementary diagram of the similarity calculation procedure, where parts (a) and (b) show the observed XRD data $\mathcal{D}$ and crystal structure factor $\mathcal{F}_k$, respectively. We resampled the data points close to $p_{m}^{(k)}$ of $\mathcal{F}_k$ from the observed data $\mathcal{D}$. The resampled data points are indicated by the red points in Figure. \ref{img:method:screening-overview}(a).

The resampled data points are denoted by the vector $\bm{y}' \in \mathbb{N}^{M_K}$. The intensity vector of the crystal structure is denoted by $\bm{I}_k=(I_{1}^{(k)}, I_{2}^{(k)}, \cdots, I_{M_k}^{(k)})^{\top} \in \mathbb{R}^{M_k}$. Our method computed the similarity between the vectors $\bm{y}'$ and $\bm{I}_k$ for each crystal structure $k$. In this study, we used cosine similarity as the vector similarity. 

\section{Scope and Limitations}\label{sec:limitations}
This section presents the two limitations of the proposed method.
\begin{itemize}
    \item The proposed method cannot refine the structural parameters owing to only the crystal structure selection and profile parameters being used as probability variables. Therefore, for precise crystal structure analysis, Rietveld analysis\cite{1967RietveldAnalysis,1969RietveldAnalysis,1993structureDetermination,2014WinCSD} must be performed with reference to the posterior distribution of the selected crystal structure and profile parameters.
    \item The proposed method automatically selects the crystal structure contained in the measurement sample from the candidate crystal structures. Therefore, crystal structures that are not included in the candidates or unknown crystal structures cannot be analyzed.
\end{itemize}

\section{Configuration}\label{sec:configuration}
\subsection{Configuration of prior distribution}
\quad We set the prior distribution over the parameter set $\Theta_\mathrm{S}$ of the profile function as follows:
\begin{eqnarray*}
    h_k &\sim& \mathcal{G}\left(k_G=4.00, \theta_G=\frac{y_\mathrm{min}-y_\mathrm{max}}{4}\right),\\
    \mu_k &\sim& \mathcal{N}(\mu_N=0.00, \sigma_N=0.05),\\
    \alpha_k &\sim& \mathcal{G}(k_G=5.00, \theta_{\alpha}=0.25),\\
    r_k &\sim& \mathcal{U}(u_U=0.00, l_U=1.00),\\
    u_k &\sim& \mathcal{G}(k_G=1.00, \theta_G=0.10),\\
    v_k &\sim& \mathcal{G}(k_G=1.00, \theta_G=0.10),\\
    w_k &\sim& \mathcal{G}(k_G=2.00, \theta_G=0.05),\\
    s_k &\sim& \mathcal{G}(k_G=2.00, \theta_G=0.05),\\
    t_k &\sim& \mathcal{G}(k_G=1.00, \theta_G=0.10).
\end{eqnarray*}
In addition, we set the prior distribution of the background parameter $\Theta_\mathrm{B}$ as follows:
\begin{eqnarray*}
    a &\sim& \mathcal{G}(k_G=2.00, \theta_G=y_\mathrm{max}),\\
    \sigma_{bg} &\sim& \mathcal{G}(k_{\sigma}=2.00, \theta_{\sigma}=2.50),\\
    r_{bg} &\sim& \mathcal{U}(u_U=0.00, l_U=1.00),\\
    b &\sim& \mathcal{U}\left(u_U=y_\mathrm{min}-\frac{\sqrt{y_\mathrm{min}}}{2}, l_U=y_\mathrm{min}+\frac{\sqrt{y_\mathrm{min}}}{2}\right).
\end{eqnarray*}
where the probability distribution $\mathcal{G}(k_G, \theta_G)$ is the gamma distribution and $k_G \in \mathbb{R}^{+}$ and $\theta_G \in \mathbb{R}^{+}$ are the shape and scale parameters, respectively. The probability distribution $\mathcal{N}(\mu_N, \sigma_N)$ is a normal distribution, and $\mu_N \in \mathbb{R}$ and $\sigma_N \in \mathbb{R}^{+}$ are the mean and standard deviation, respectively. Whereas, the probability distribution $\mathcal{U}(u_U, l_U)$ is a uniform distribution, with $u_U \in \mathbb{R}$ and $l_U \in \mathbb{R}$ being the maximum and minimum values, respectively. Further, the values $y_\mathrm{min} \in \mathbb{N}, y_\mathrm{max} \in \mathbb{N}$ are $y_\mathrm{min}=\mathrm{min}(\bm{y})$ and $y_\mathrm{max}=\mathrm{max}(\bm{y})$, where $\bm{y} = (y_1, y_2, ..., y_N)^{\top}$.

\subsection{Configuration of the sampling algorithm}
\quad For the exchange MC simulation, we performed 1000 steps of calculations and rejected 1000 of them as burn-in. The inverse temperature was set as follows:
\begin{eqnarray}
    \beta_{\tau} &=& \left\{
        \begin{array}{ll}
        0 & (\tau = 0)\\
        \eta^{\tau-T} & (\tau \neq 0),
        \end{array}
        \right.\\
    \text{where} \:\: \tau &\in& \{0, 1, 2, ..., T\} ,
\end{eqnarray}
where the proportion $\eta \in \mathbb{R}^{+}$ was set to $\eta=1.2$, and the number of temperatures $T \in \mathbb{N}$ was set to $T=64$. The exchange of parameter sets between replicates was performed at each step.

\subsection{Calculator Specification}
\quad The calculator specifications were AMD Ryzen Thread ripper 3990X (64 core, 128 thread), 256GB DDR4-3200/PC4-25600SD, Ubuntu 18.04.5 LTS. We performed sampling using the REMC method with 32 threads.

\subsection{Configuration of candidate crystal structures}\label{sec:candidate}
\quad We prepared 50 candidate crystal structures from the AtomWork\cite{AtomWork}, which is an inorganic material database containing data on the crystal structures, X-ray diffraction, properties, and state diagrams of inorganic materials extracted from scientific and technical literature. We selected 50 candidates based on the condition that they contained titanium (Ti) or oxygen (O) in their composition because this study analyzed the XRD data of the titanium dioxide $\rm TiO_2$ samples. Table \ref{table:candidate-crystal-structures} lists the 50 prepared candidate crystal structures.

\begin{table}
    \centering
    \caption{Fifty candidate crystal structures prepared from the AtomWork\cite{AtomWork}, which is the inorganic material database.}
    \label{table:candidate-crystal-structures}
    \scalebox{1.2}[1.2]{
    \begin{tabular}{ccc|cc}
        \hline
        &  & state & chemical & crystal \\
        & composition & structure & composition & structure \\
        \hline \hline
        01 & $\mathrm{TiO_{2}}$      & $\mathrm{Rutile}$         & $\mathrm{TiO_{2}}$           & $\mathrm{Anatase}$ \\
        02 & $\mathrm{TiO_{2}}$      & $\mathrm{Brookite}$       & $\mathrm{O_{2}}$             & $\mathrm{O_2}$ \\
        03 & $\mathrm{Ti_{3}O_{5}}$  & $\mathrm{Ta_{3}N_{5}}$    & $\mathrm{Ti}$                & $\mathrm{Mg}$ \\
        04 & $\mathrm{Ti}$           & $\mathrm{W}$              & $\mathrm{TiO_{2}}$           & $\mathrm{Fe_{2}N_{0.94}}$ \\
        05 & $\mathrm{Ti_{2}O_{5}}$  & $\mathrm{a}$              & $\mathrm{TiO_{2}}$           & $\mathrm{CdI_{2}}$ \\
        06 & $\mathrm{TiO_{2}}$      & $\mathrm{Al_{2}O_{3}}$    & $\mathrm{TiO_{0.2}}$         & $\mathrm{Mg}$ \\
        07 & $\mathrm{Ti_{5}O_{9}}$  & $\mathrm{Ti_{5}O_{9}}$    & $\mathrm{Ti_{7}O_{13}}$      & $\mathrm{Ti_{7}O_{13}}$ \\
        08 & $\mathrm{Ti_{9}O_{17}}$ & $\mathrm{Ti_{9}O_{17}}$   & $\mathrm{Ti_{4}O_{7}}$       & $\mathrm{a}$ \\
        09 & $\mathrm{Ti_{4}O_{7}}$  & $\mathrm{Ti_{4}O_{7}}$    & $\mathrm{Ti_{4}O_{7}}$       & $\mathrm{b}$ \\
        10 & $\mathrm{TiO_{2}}$      & $\mathrm{Fe_{2}N_{0.94}}$ & $\mathrm{TiO}$               & $\mathrm{NaCl}$ \\
        11 & $\mathrm{Ti_{4}O_{5}}$  & $\mathrm{Ti_{4}O_{5}}$    & $\mathrm{TiO_{2}}$           & $\mathrm{CdI_{2}}$ \\
        12 & $\mathrm{Ti_{4}O_{7}}$  & $\mathrm{Ti_{4}O_{7}}$    & $\mathrm{Ti_{4}O_{7}}$       & $\mathrm{a}$ \\
        13 & $\mathrm{Ti_{4}O_{7}}$  & $\mathrm{b}$              & $\mathrm{Ti_{5}O_{9}}$       & $\mathrm{Ti_{5}O_{9}}$ \\
        14 & $\mathrm{Ti_{9}O_{17}}$ & $\mathrm{Ti_{9}O_{17}}$   & $\mathrm{Ti_{6}O_{11}}$      & $\mathrm{Ti_{6}O_{11}}$ \\
        15 & $\mathrm{Ti_{7}O_{13}}$ & $\mathrm{Ti_{7}O_{13}}$   & $\mathrm{Ti_{8}O_{15}}$      & $\mathrm{Ti_{8}O_{15}}$ \\
        16 & $\mathrm{Ti_{6}O_{11}}$ & $\mathrm{Ti_{6}O_{11}}$   & $\mathrm{Ti_{3}O}$           & $\mathrm{Ti_{3}O}$ \\
        17 & $\mathrm{TiO}$          & $\mathrm{TiO}$            & $\mathrm{Ti_{0.84}O_{0.84}}$ & $\mathrm{TiO}$ \\
        18 & $\mathrm{Ti_{4}O_{5}}$  & $\mathrm{Ti_{4}O_{5}}$    & $\mathrm{Ti}$                & $\mathrm{Ti}$ \\
        19 & $\mathrm{Ti_{6}O}$      & $\mathrm{Ti_{6}O}$        & $\mathrm{Ti_{6}O}$           & $\mathrm{Ti_{6}O}$ \\
        20 & $\mathrm{TiO_{2}}$      & $\mathrm{Mg}$             & $\mathrm{TiO_{2}}$           & $\mathrm{ZrO_{2}}$-b \\
        21 & $\mathrm{TiO_{2}}$      & $\mathrm{MnO_{2}}$        & $\mathrm{Ti_{2}O_{5}}$       & $\mathrm{b}$ \\
        22 & $\mathrm{Ti_{3}O_{5}}$  & $\mathrm{V_{3}O_{5}}$     & $\mathrm{TiO}$               & $\mathrm{WC}$ \\
        23 & $\mathrm{TiO_{2}}$      & $\mathrm{VO_{2}}$-b       & $\mathrm{TiO_{2}}$           & $\mathrm{MnO_{2}}$ \\
        24 & $\mathrm{Ti_{2}O_{5}}$  & $\mathrm{b}$              & $\mathrm{Ti_{2}O_{5}}$       & $\mathrm{a}$ \\
        25 & $\mathrm{TiO_{2}}$      & $\mathrm{VO_{2}}$-b       & $\mathrm{Ti_{3}O_{5}}$       & $\mathrm{V_{3}O_{5}}$ \\
        \hline
    \end{tabular}
    }
\end{table}

\section{Results and discussion}
\subsection{Fitting results in actual measurement data}\label{sec:experiment:001}
\quad We conducted a calculation experiment on the measured XRD data. The measurement sample was a mixture of multiple types of $\rm TiO_2$: Anatase, Brookite, and Rutile. The mixture ratios were equal (1/1/1 wt. \%). We prepared measurement samples such that the crystalline phases were homogeneous. Consequently, we measured the XRD data by using monochromatic X-rays of $\rm{Cu} \ \rm{K}_{\alpha1}$. Further, a non-reflecting plate cut from a specific orientation of a single crystal of silicon was used as the sample plate. The diffraction angles $2\theta$ were in the range of 10--60[$^{\circ}$], with $2\theta$ of $\bm x = (10.00, 10.02, 10.04, ..., 60.00)^{\top}$.

\begin{figure}
  \centering
  \includegraphics[width=\linewidth]{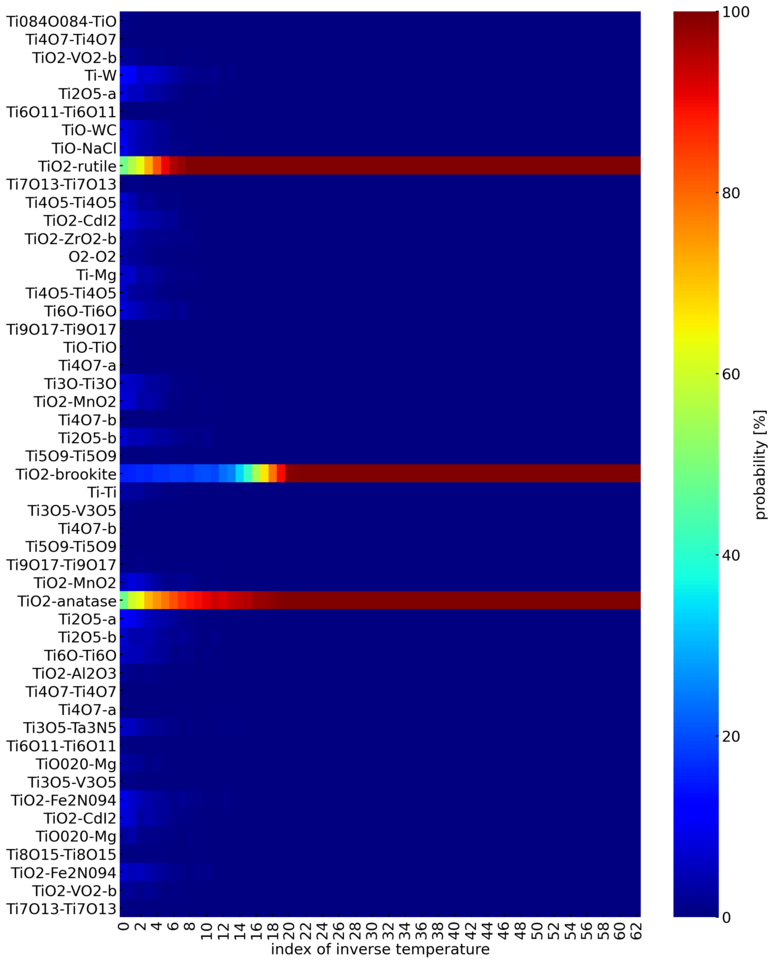}
  \caption{Selection results from 50 candidates for each temperature in the REMC method. The x- and the y-axes denote the candidate crystal structures and index of inverse temperature $\tau$, respectively. This figure shows a visualization of the indicator probability $P(\bm{g}|Y;\beta=\beta_{\tau})$ [\%]. The large index corresponds to lower temperatures. The color scale denotes the sampling frequency of $g_k=1$ on a log scale. The dark red indicates the presence of crystal structure in the measured sample.}
  \label{img:obs:heatmap-temp-indicator-full}
\end{figure}

Figure \ref{img:obs:heatmap-temp-indicator-full} presents the selection results for each temperature obtained using the REMC method. The x- and y-axes denote the candidate crystal structures and inverse temperature index $\tau$, respectively. This figure visualizes the probability of indicator $P(\bm{g}|Y;\beta=\beta_{\tau})$ [\%]. The proposed method estimated the crystal structure of a sample from 50 candidates. Candidate crystal structures were obtained from AtomWork as described in Section \ref{sec:candidate}. A high index corresponds to a lower temperature. The color scale indicates the probability of $g_k=1$ calculated from the sampling frequency. The dark red color indicates the presence of a crystal structure in the measured sample. The result for the lowest temperature ($\tau = 64$) shows that our method could select the true crystal structures, that is, Anatase, Brookite, and Rutile, with 100 [\%] probability. A computational time of approximately 3 h was required to obtain this result. Therefore, our method can be used to estimate the crystal structures of a sample by analyzing the full diffraction profile using Bayesian inference. Thus, the contribution of our method is the simultaneous identification of profile parameters and crystal structures and the provision of their posterior distributions. 

As shown in Figure \ref{img:obs:heatmap-temp-indicator-full}, the selection probability of Brookite decreases at medium to high temperatures compared to those of Anatase and Rutile. This suggests that Brookite was more difficult to identify than Anatase and Rutile. In crystallography, Brookite is a low-temperature phase and is known to exhibit a poorer crystal structure than Rutile, which is high-temperature-stable. This difficulty in its determination is believed to originate from the low crystallinity of Brookite.

\begin{figure}
  \centering
  \includegraphics[width=\linewidth]{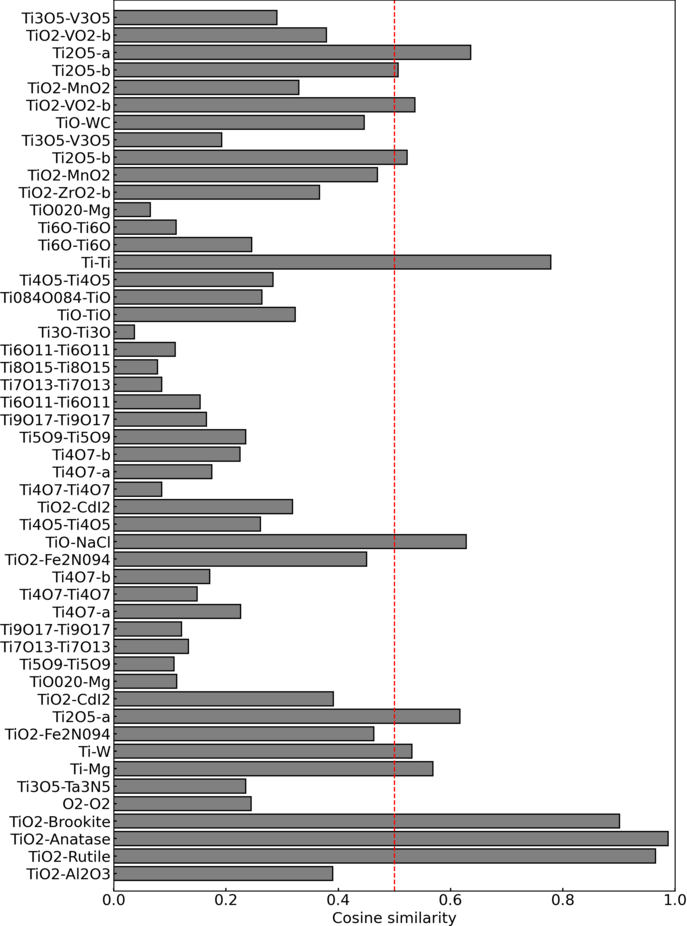}
  \caption{Cosine similarity between the measurement XRD data $\mathcal{D}$ and the crystal structure factors $\mathcal{F}$ for prescreening. In this figure, the red line denotes the threshold value set at 0.5.}
  \label{img:obs:cosine_similarity}
\end{figure}

An analysis using all 50 candidates would require a considerable amount of time. Therefore, we performed prescreening using the cosine similarity described in Section \ref{sec:technique}. Figure \ref{img:obs:cosine_similarity} shows the cosine similarity between the measured XRD data $\mathcal{D}$ and crystal structure factors $\mathcal{F}$ during prescreening. In this figure, the red line denotes the threshold value, which was set to 0.5. The y-axis denotes cosine similarity. We performed the analysis using crystal structure factors with a cosine similarity greater than 0.5. This prescreening narrowed the list from 50 to 12 candidates. This is expected to result in significant reduction in the computational costs.

\begin{figure}
  \centering
  \includegraphics[width=\linewidth]{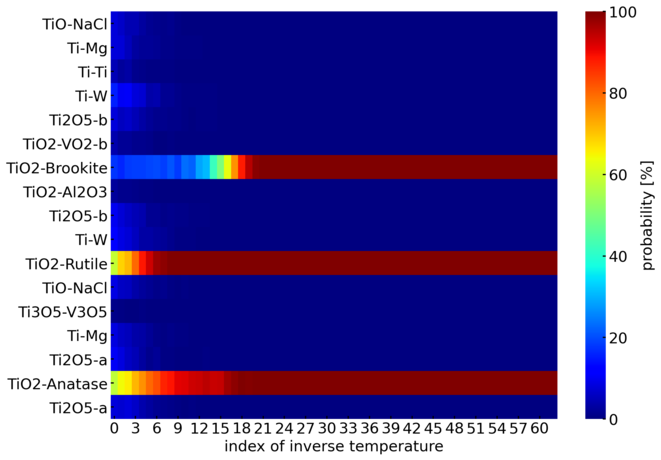}
  \caption{Selection result from 12 candidates for each temperature in the REMC method. The x- and the y-axes denote the candidate crystal structures and the index of the inverse temperature $\tau$. This figure shows a visualization of the indicator probability $P(\bm{g}|Y;\beta=\beta_{\tau})$ [\%]. The large index corresponds to lower temperatures. The color scale denotes the sampling frequency of $g_k=1$ on a log scale. The dark red color indicates the presence of crystal structure in the measured sample.}
  \label{img:obs:heatmap-temp-indicator}
\end{figure}

We analyzed the measured XRD data using the 12 candidates that were narrowed down by prescreening. Figure \ref{img:obs:heatmap-temp-indicator} presents the selection results for each temperature obtained using the REMC method. The x- and y-axes denote the candidate crystal structures and the index of the inverse temperature $\tau$. The proposed method could select the true crystal structures of Anatase, Brookite, and Rutile with 100 [\%] probability. The crystal structures and results of sampling all 50 candidates were successfully identified (shown in Figure \ref{img:obs:heatmap-temp-indicator-full}). The computation required approximately 1 h, and pre-screening reduced the computational cost by a factor of three. These results indicate that prescreening can effectively improve the efficiency of the calculations. However, prescreening may exclude true crystal structures from the candidates.

\begin{figure}
  \centering
  \includegraphics[width=\linewidth]{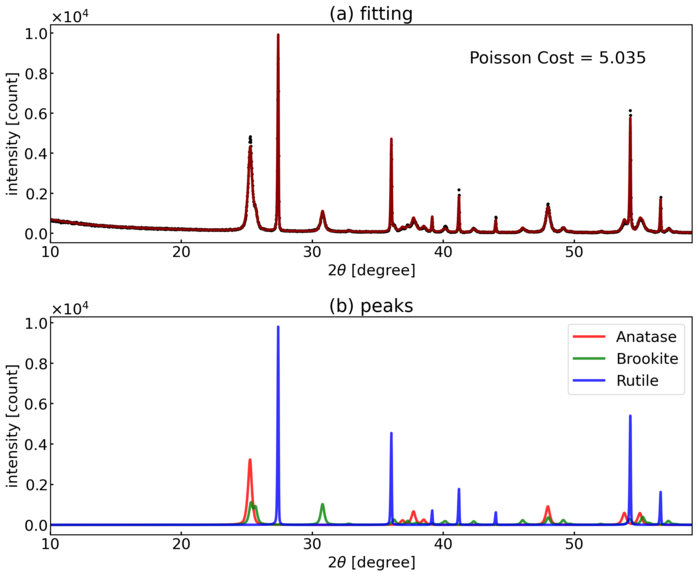}
  \caption{Result of profile analysis in the measurement XRD data using our method [(a): Fitting result via profile function in the measurement XRD data. In this figure, the black and the red lines indicate the measurement XRD data and the fitting profile functions, respectively. (b): Peak components in three crystal structures of $\rm TiO_2$; Anatase, Brookite, and Rutile. The red, green, and blue lines indicate the peaks of Anatase, Brookite, and Rutile, respectively.]}
  \label{img:obs:fitting}
\end{figure}

Figure \ref{img:obs:fitting}(a) shows the fitting results via the profile function in the measurement XRD data. In Figure \ref{img:obs:fitting}(a), the black and red lines indicate the measured XRD data and the fitting profile functions, respectively. Figure \ref{img:obs:fitting}(b) shows the peak components of the three crystal structures of $\rm TiO_2$: Anatase, Brookite, and Rutile. The red, green, and blue lines indicate the peaks of Anatase, Brookite, and Rutile, respectively. As shown in this figure, the estimated profile function faciliated a good fit of the XRD data. The mean Poisson cost $E(\hat{\Theta})$ was 5.026.


\begin{figure}
    \centering
    \includegraphics[width=0.7\linewidth]{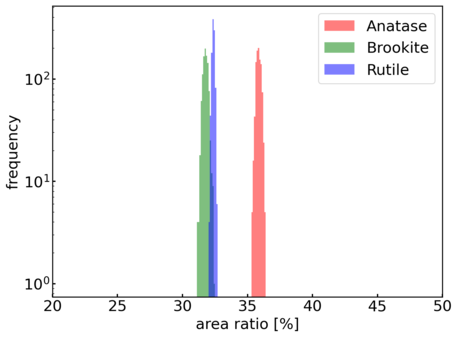}
    \caption{Expanded view of the posterior distribution of the peak area ratio when analyzing the measurement XRD data using the proposed method. The units for the axes are percentages [\%].}
    \label{img:obs:post-dist-ratio}
\end{figure}


Figure \ref{img:obs:post-dist-ratio} shows an expanded view of the posterior distribution of the peak used to determine its shape of the posterior distribution. In Figure. \ref{img:obs:post-dist-ratio}, the red, green, and blue histograms correspond to the posterior distributions of Anatase, Brookite, and Rutile, respectively. As evident, the posterior distribution of Rutile, which has the best crystallinity, exhibited a sharper shape than Anatase and Brookite. The shape of the posterior distribution was similar to that of a quadratic function, where the y-axis represents a logarithmic scale. This implies that the posterior distribution exhibits a Gaussian probability distribution shape. The MAP estimate of the ratio was $\mathrm{Anatase} : \mathrm{Brookite}: \mathrm{Rutile} = 35.7 : 31.8 : 32.5$ [\%]. Because the structural ratio of the preparation is $\mathrm{Anatase} : \mathrm{Brookite}: \mathrm{Rutile} = 33.3\dot{3} : 33.3\dot{3} : 33.3\dot{3}$ [\%], the proposed method is considered a reasonable estimation.

\begin{figure}
  \centering
  \includegraphics[width=\linewidth]{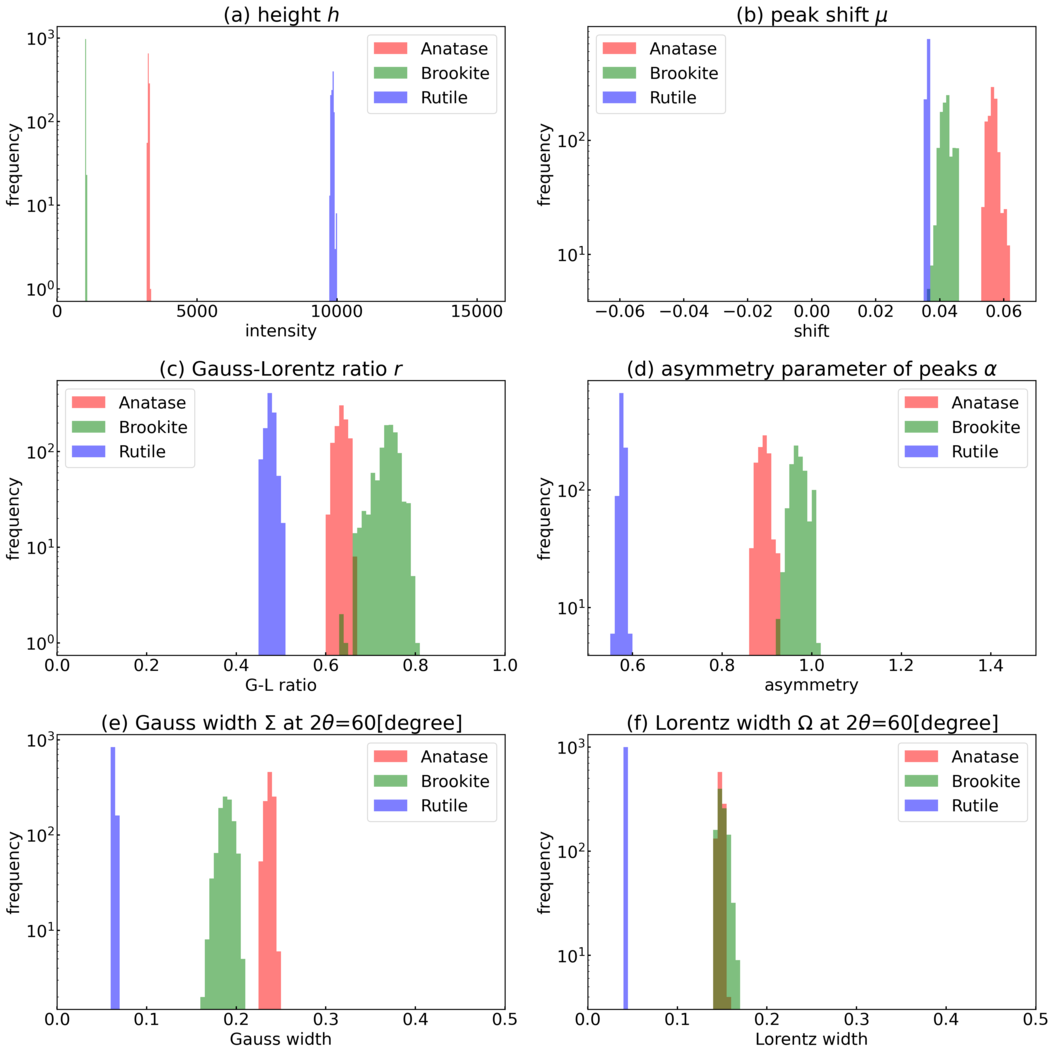}
  \caption{Posterior distribution of the profile parameters in the measurement XRD data. The red, green, and blue histograms correspond to the posterior distribution of Anatase, Brookite, and Rutile, respectively. [(a): Peak height $h$, (b): peak shift $\mu$, (c): Gauss-Lorentz ratio $r$, and (d): asymmetry parameter $\alpha$]. (e) and (f) are Gauss width $\Sigma(x_i;u_k, v_k, w_k, \alpha_{k})$ and Lorentz width $\Omega(x_i;u_k, v_k, w_k, \alpha_{k})$, where $x_i$ is $2\theta = 60$ [$^{\circ}$]. The black dot-dash line is a true parameter of the profile function.}
  \label{img:obs:post-dist-param}
\end{figure}

Figure \ref{img:obs:post-dist-param} shows the posterior distribution of the profile parameters when analyzing the measurement XRD data using the proposed method. The red, green, and blue histograms represent the posterior distributions of Anatase, Brookite, and Rutile, respectively. Figure \ref{img:obs:post-dist-param}(a)--(d) show the peak height $h$, peak shift $\mu$, Gaussian–Lorentz ratio $r$, and asymmetry parameter $\alpha$, respectively. Figure \ref{img:obs:post-dist-param}(e) and (f) show the Gaussian width $\Sigma(x_i;u_k, v_k, w_k, \alpha_{k})$ and the Lorentz width $\Omega(x_i;u_k, v_k, w_k, \alpha_{k})$, where $x_i$ is $2\theta = 60$ [°]. As indicated in part (a) of this figure, the height $\bm h$ can be estimated with high precision using the proposed method. The figure shows that the peak shifts for all three crystal structures were positive ($\mu=0.04 \sim 0.06$). This may be attributed to minute calibration deviations in the measurement device such as eccentricity and zero-point errors. As shown in parts (e) and (f) of this figure, the peak width of Rutile was narrow, indicating good crystallinity. Furthermore, we confirmed that Rutile with good crystallinity exhibited a sharp posterior distribution shape for most of the profile parameters. By contrast, Brookite with poor crystallinity, tended to exhibit a broad posterior distribution. This indicates that a structure with good crystallinity provides a highly precise estimation.

\section{Conclusion}
\quad The knowledge of the probability that a sample contains a candidate crystal structure from full-range XRD data considering both the diffraction angles of the peaks and the profile functions, is essential. This study aimed at the Bayesian estimation of the structure contained in a sample from a large number of crystal structure candidates in the analysis of XRD data. Therefore, indicator vectors were introduced into the profile function and the XRD data were analyzed by sampling the posterior distribution using the REMC method. Consequently, we succeeded in identifying the true crystal structures of 50 candidates with high probability. The proposed method also estimated the mixing ratio of the selected crystal structures with high precision. In this study, we provide reasonable results that allow clearer identification of the crystal structure for more crystalline structures. Our method is a highly sensitive and probabilistic analysis method that can automatically identify crystal structures from full-range XRD data.



\section*{Acknowledgment}
This work was supported by MEXT KAKENHI under grant (number 18K05191); and JSPS KAKENHI under grant (number 19K12154). 

\bibliographystyle{unsrt} 
\bibliography{main} %

\appendix

\end{document}